\documentclass[12pt]{article}
\usepackage{graphicx}
\usepackage{caption}
\usepackage{amsmath}
\usepackage{amssymb}
\usepackage{mathtools}
\usepackage[letterpaper,top=2cm,bottom=2cm,left=3cm,right=3cm,marginparwidth=1.75cm]{geometry}
\usepackage[colorlinks=true, allcolors=blue]{hyperref}
\usepackage{xcolor}
\usepackage[mathlines]{lineno}
\usepackage{authblk}
\usepackage{bm}
\usepackage{lineno}
\usepackage{float}

\usepackage{titlesec}
\date{}

\usepackage[sort&compress, numbers, round]{natbib}
\usepackage{url}

\newcommand{\upcite}[1]{\textsuperscript{\cite{#1}}}

\title{Revealing dynamics of non-autonomous complex systems from data}

\author[1,2]{\normalsize Chengzuo Zhuge}
\author[2,3]{\normalsize Zheng Jiang}
\author[4]{\normalsize Zhefan Xu}
\author[5,*]{\normalsize Wei Chen}

\affil[1]{\footnotesize School of Mathematical Sciences, Beihang University, Beijing, 100191, China}
\affil[2]{\footnotesize Key Laboratory of Mathematics, Informatics and Behavioral Semantics (LMIB), Beihang University, Beijing, 100191, China}
\affil[3]{\footnotesize School of Artificial Intelligence, Beihang University, Beijing, 100191, China}
\affil[4]{\footnotesize Department of Mechanical Engineering, Carnegie Mellon University, 5000 Forbes Ave, Pittsburgh, PA, 15213, USA}
\affil[5]{\footnotesize School of Systems Science, Beijing Normal University, Beijing, 100875, China}
\affil[*]{\footnotesize Corresponding author: chwei@bnu.edu.cn}

\begin{document}
\maketitle

\begin{abstract}
Discovering governing equations from data is crucial for understanding complex systems in many diverse fields from science to engineering. Yet, there still is a lack of versatile computational toolbox to deal with this long standing challenge due to the inherent non-autonomicity and unknowability of the underlying dynamics. Here, we introduce a data-driven approach for inferring non-autonomous dynamical equations by identifying an optimal set of basis functions within the model space, enabling the reconstruction of complex systems behavior under simplified prior specifications. Our method demonstrates effectiveness in equation discovery on canonical synthetic systems such as cusp bifurcation and coupled Kuramoto oscillators. Furthermore, we extend the application of this approach to leaf cellular energy, unmanned aerial vehicle navigation, chick-heart aggregates, and marine fish community under simple basis function libraries. Leveraging the inferred equations, we accurately predict the evolution of these empirical systems and further uncover their governing laws. Our approach offers a novel paradigm to reveal the underlying dynamics of a wide range of real-world systems.
\end{abstract}

\section{Introduction}
From cellular metabolism\upcite{RN263,RN249}, to animal population migration\upcite{RN253,RN254}, to autonomous driving of vehicles\upcite{RN248,RN252}, the intricate evolution of numerous real-world systems is fundamentally governed by unknown system dynamics. Given the critical importance of the governing equations underlying a system's dynamical behavior for understanding and predicting its evolution, several advanced methodologies have been proposed in recent years to uncover the dynamical equations from observed data\upcite{RN216,RN306,RN257,RN211,RN213,RN245,RN243,RN246,RN292,RN215}. These methods typically define a model space that potentially contains the system's latent dynamics through the linear combinations of predefined basis functions. Then the optimal combination in this model space is learned from observed data to capture the governing dynamics, for example, by applying sparse regression to find parsimonious model\upcite{RN216}, or by integrating local fine-tuning with global regression to derive compact equations\upcite{RN243}.

Despite these advancements, most existing methodologies have focused on autonomous systems, there still is a lack of systematic investigation of non-autonomous systems, particularly those involving network topologies\upcite{RN204,RN295,RN296,RN297,RN280,RN214}. However, real-world systems are predominantly non-autonomous, with the evolution of their states often being subject to external time-varying parameters\upcite{RN100}. For instance, population dynamics are not only governed by internal interactions such as mutualism and competition, but also driven by external factors like temperature fluctuations\upcite{RN316,RN317}. Such non-autonomicity leads to the emergence of particularly noteworthy dynamical behaviors, such as critical transitions\upcite{RN63,RN232} and seasonal patterns\upcite{RN299,RN303}. To extend existing methodologies to non-autonomous systems, current approaches typically augment the candidate library to include not only functions of the system's state variables but also those of the forcing parameters and their coupled terms\upcite{RN283,RN313,RN209}. However, this framework faces several fundamental challenges when applied to real-world non-autonomous dynamics. First, the forcing parameters in real-world systems often exhibit unobservability. Consequently, their values are not available for assigning to the basis functions, and thus the system's governing equations cannot be inferred in most cases. As a remedy, an intuitive strategy for modeling non-autonomous dynamics is to replace forcing parameters with a readily observable time variable within the basis functions\upcite{RN216}. Yet, the validity of such a simplification in the basis functions has not been systematically verified. It remains unclear what forms of the time variable in the basis functions are valid, while invalid basis functions can directly lead to biased dynamics inference. Second, the dynamics of real-world systems are highly complex and not fully knowable, especially for non-autonomous dynamics with intricate signal couplings. Current methods address this issue by identifying the optimal combination of basis functions from an extensive candidate library to reconstruct system dynamics\upcite{RN243,RN246}. Such a library comprises a set of basis functions whose selection is heuristic and often relies on expert knowledge. However, the search for the optimal combination of basis functions within a broad model space incurs expensive computational costs\upcite{RN243,RN246,RN292}. Meanwhile, the heuristically selected basis functions may still lack sufficient completeness to span a model space that contains the true governing equation of real-world system. As a consequence, existing methodologies struggle to extend their applicability to non-autonomous dynamics inference in real-world settings.

These challenges necessitate a framework capable of inferring dynamical equations without rigid dependence on the completeness of predefined libraries or accessibility of forcing parameters. In this work, we introduce a novel paradigm that shifts the focus from merely finding the optimal linear combination of predefined basis functions to adaptively constructing the basis functions themselves within the given model space. Specifically, we first prove a theorem for the basis functions of non-autonomous systems, which provides a theoretical guidance for developing a method to infer equations for complex system dynamics. The theorem states that, within a given model space, there exist infinitely many sets of basis functions, each parameterized by distinct driving variables, to represent the multi-parameter driven equations hidden in this space. Based on the theorem, we propose an approach for accurately inferring equations by adaptively identifying an optimal set of basis functions within a given model space. These optimal basis functions, identified through a unique driving variable detected from the observed data, introduce minimal fitting and numerical errors in the process of equation inference. Such improvements in the basis functions enable the uncovering of non-autonomous dynamics even when the forcing parameters are inaccessible, and promise accurate reconstruction of the true dynamics without requiring rigid prior knowledge of the system dynamics.

To validate our method's equation discovery capability, we employ two canonical synthetic systems: the cusp bifurcation normal form\upcite{RN66} and the coupled Kuramoto oscillators\upcite{RN260}. The governing equations obtained via the optimal basis functions achieve higher fidelity to the ground truth compared to those obtained using basis functions that depend on forcing parameters or time variable. This improvement, observed in both the recovered functional forms and coefficient estimates, remains robust as the size of the basis function library increases. Notably, even if key functional terms of the true equations are omitted in the basis function library, our method consistently recovers the system dynamics through a surrogate model, where comparable approaches fail. Moreover, when dealing with real-world systems where the true underlying equations are highly intricate and not fully knowable, our method can still accurately infer representations of the underlying dynamics under simple basis function libraries. We demonstrate this capability on non-autonomous complex systems from diverse real-world domains, including leaf cellular energy status\upcite{RN249}, unmanned aerial vehicle (UAV) navigation, spontaneously beating chick-heart aggregates\upcite{RN108}, and natural marine fish community\upcite{RN204}. Leveraging the inferred equations, we not only generate accurate predictions for the evolution of these empirical systems, but our further analysis also indicates that the equations essentially reveal their underlying governing laws. Specifically, we provide empirical evidence that critical transitions in nature, such as the cellular energy collapse under hypoxia and the arrhythmia of chick-heart aggregates induced by drugs, can be understood through bifurcation theory\upcite{RN66}. Moreover, we reveal the mechanisms underlying population fluctuations in marine fish community. Interestingly, we further find that the control algorithms of UAV can be distilled from its trajectory data in the form of dynamical equations. These results imply that our methodology can offer promising insights for understanding the dynamics of complex real-world systems.

\section{Results}
\subsection{Overview of optimal basis inference framework} 
Consider a non-autonomous complex system which is governed by the equation
\begin{equation*}
    \dot{\bm x}=f(\bm x,\bm{\Phi}),
\end{equation*}
where $\bm x \in \mathbb{R}^d$ is the system's state, and $\bm{\Phi} \in \mathbb{R}^n$ are the external time-varying parameters that drive the evolution of the system. Given the state time series $\{\mathbf{x}_i\}_{i=1}^T$ sampled at discrete times $i=1,\dots,T$, the observed temporal variation of $\bm{\Phi}$ can be characterized through a sign sequence $\{s_i\}_{i=1}^{T-1}$, where $s_i \in \{+1,-1\}$ indicates whether $\bm{\Phi}$ increases or decreases at each time step. Specifically, $\bm{\Phi}_{i+1}=\bm{\Phi}_{i}+s_i\Delta \bm{\Phi}$, where $\Delta \bm{\Phi}$ is a fixed increment. The equation to be inferred lies in a potential model space spanned by basis functions of the system's state variables $\bm{x}$ and parameters $\bm{\Phi}$ (Fig. 1a), where each basis function is obtained as a finite product of elements from the sub-basis $\mathcal{B}$ (see details in Methods). In this work, we prove that the dynamical equation can be represented by another set of basis functions involving a variable $\nu$, whose value at $i$-th time point is determined by $\nu_{i+1}=\nu_i+s_i\Delta \nu$. The theorem holds for an arbitrary initial value $\nu_1$ and an arbitrary constant step size $\Delta\nu$ (see details for this theorem in Methods). This indicates that this set of basis functions spans the same model space as the original basis functions with time-varying parameters $\bm{\Phi}$. Here, we state that the set of basis functions with the variable $\nu$ is equivalent to that with parameters $\bm{\Phi}$ (Fig. 1b).

After constructing the set of basis functions with variable $\nu$, we can proceed to infer the governing equations of the system dynamics based on the time-series state data $\mathbf{X}=\{\mathbf{x}_i\}_{i=1}^T$, without requiring observation of $\bm{\Phi}$. Specifically, we infer the governing equation by solving the optimal coefficient matrix $\mathbf{A}_{\nu}$ from $\dot{\mathbf{X}}=\Theta_{\nu}\mathbf{A}_{\nu}$, where $\dot{\mathbf{X}}$ is the finite difference approximation of the derivative of $\mathbf{X}$\upcite{RN244}, and $\Theta_{\nu}$ is the feature matrix constructed by incorporating both $\mathbf{X}$ and the time-series data of $\nu$ into these basis functions. The theorem further demonstrates that the inferred equations, obtained through solving $\dot{\mathbf{X}}=\Theta_{\nu}\mathbf{A}_{\nu}$, are theoretically equivalent for all $\nu$, indicating that the underlying dynamics they govern are identical despite differences in their forms.

In practice, however, numerical errors introduced by the pseudo-inverse operation in solving $\mathbf{A}_{\nu}$ can significantly bias the estimated coefficients, particularly when inferring non-autonomous dynamics featuring complex driving signal patterns. Therefore, it is crucial to develop strategies to reduce pseudo-inverse errors and thereby improve the reliability of our inferences. Specifically, our framework identifies a unique optimal driving variable among all equivalent alternatives that minimizes the inference inaccuracy amplified by numerical errors. We employ the grid search in conjunction with the numerical error adjusted Akaike's information criterion ($\varepsilon$AIC) to obtain the optimal pair $(\nu_1,\Delta \nu)$, which finds a set of basis functions that balance numerical accuracy, fitting accuracy, and model complexity, as shown in Figs. 1c-e. Essentially, given the posterior observed data, our framework finds a set of basis functions that optimally represent the latent dynamical equation in the model space, thereby enabling the forecasting of the system's evolution (Fig. 1f).

\begin{figure}[htbp!]
\centering
\includegraphics[width=\textwidth]{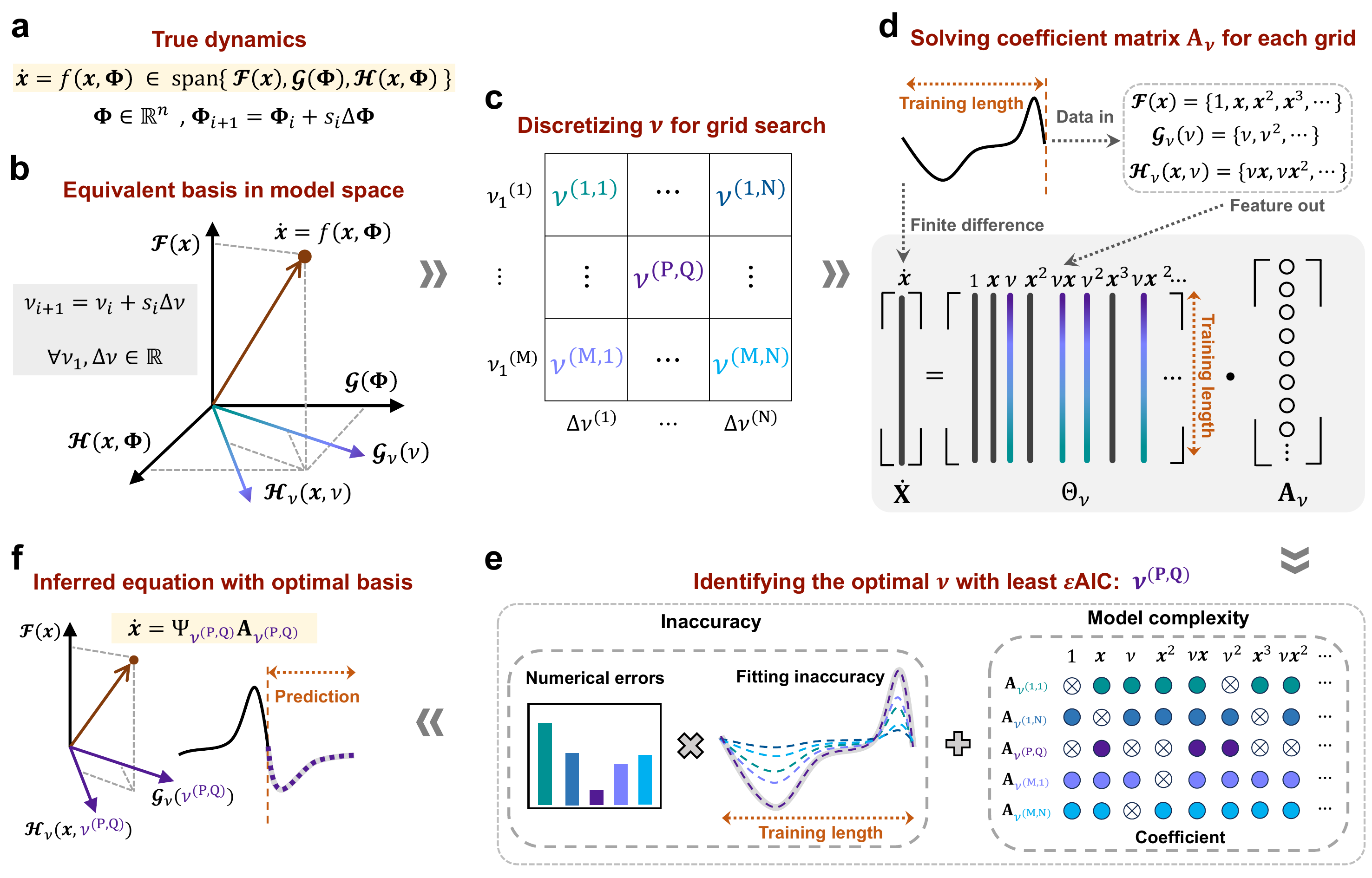}
\caption*{\fontsize{10}{12}\selectfont \textbf{Fig.1 \textbar\ Overview of optimal basis inference framework.} (\textbf{a}). The system dynamics are captured by the unknown equation $f(\bm{x},\bm{\Phi})$ within a potential model space, which we aim to reconstruct from data. (\textbf{b}). Constructing an equivalent set of basis functions with variable $\nu$ in the model space to represent $f(\bm{x},\bm{\Phi})$, which remains theoretically equivalent when solved from data. (\textbf{c}-\textbf{e}). Grid search is used to identify the optimal variable $\nu$ for equation inference: (\textbf{c}) discretization of the initial value $\nu_1$ and step size $\Delta \nu$ for an $\text{M}\times\text{N}$ grid search, with $\text{M}=11$ and $\text{N}=12$ in our experiments; (\textbf{d}) solving the coefficient matrix $\mathbf{A}_{\nu}$ for each pair $(\nu_1,\Delta \nu)$, and the resulting equation is $\dot{\bm{x}}=\Psi_{\nu}\mathbf{A}_{\nu}$; (\textbf{e}) finding the optimal driving variable with least $\varepsilon$AIC. (\textbf{f}). Predicting state evolution based on the equation inferred by the optimal basis functions within the model space.}
\end{figure}

\subsection{Inferring dynamics of synthetic systems}
To validate the capability of discovering governing equations from data, we applied our approach to the normal form of cusp bifurcation\upcite{RN66} and the coupled Kuramoto oscillators\upcite{RN260}, which are representative synthetic systems driven by multiple forcing parameters and by a single forcing parameter, respectively. Given the sparsity of dynamics in the synthetic systems, we employed LASSO regression\upcite{RN216} within the optimal basis inference framework to uncover their governing equations.

Cusp bifurcation is a canonical dynamical system structure explaining various natural phenomena from critical transitions\upcite{RN63} to the operation of simple machines\upcite{RN301,RN300}. The normal form of cusp bifurcation is given by
\begin{equation*}
    \dot{x}=\phi_1+\phi_2x-x^3,
\end{equation*}
where $\phi_1$ and $\phi_2$ are the bifurcation parameters. The cusp bifurcations occur at a bifurcation curve $\{(\phi_1,\phi_2):4\phi_2^2-27\phi_1^3=0\}$, where the system shifts from one state to another. Based on this equation, we generated 100 time series by varying the bifurcation parameters, each simulated from equilibrium with different initial values of $(\phi_1,\phi_2)$. Specifically, the bifurcation parameters $(\phi_1,\phi_2)$ were varied linearly from each point of $10\times10$ grid ($[0.2,0.4,\dots,1.8,2]\times[4,4.5,\dots,8,8.5]$) to the terminal point $(3,3)$, with an example shown in Fig. 2a. Each time series consists of 1000 data points, with a sampling interval of 0.01, and the first 500 points are used for training. Then, we tested our method using the basis functions constructed from the sub-basis $\mathcal{B}=\{\bm{x},\nu\}$ (degree $k=3$) on these time series, where we set $\nu$ as a monotonically varying variable (i.e., $s_i\equiv +1$) to match the monotonic variation of $(\phi_1,\phi_2)$. For the case shown in Fig. 2a, we obtained the inferred equation based on the optimal set of basis functions identified through the grid search, which is consistent with the true equation (Fig. 2b). For all 100 time series, our method faithfully recovered the coefficients and functional terms of the true equations, allowing the inferred equations to consistently achieve high accuracy in predicting the dynamics (Fig. 2c). These results demonstrate that our method successfully uncovers the governing equations of the cusp bifurcation.

The Kuramoto model characterizes the dynamical behavior of a class of complex systems exhibiting collective synchronization phenomena through coupled phase oscillators. The dynamics are governed by the following equations
\begin{equation*}
\dot{\theta_i}=\omega_i+\sigma\sum\limits_{j=1}^{N}A_{ij}\sin(\theta_j-\theta_i),
\end{equation*}
where $\theta_i$ and $\omega_i$ denote the phase and the natural frequency of the $i$-th oscillator, respectively. $\sigma$ represents the coupling strength of the interactions, $A=(A_{ij})$ is the adjacency matrix of the network. The dynamics of coupled Kuramoto oscillators depend significantly on the coupling strength $\sigma$. Specifically, when the coupling strength $\sigma$ is small, the oscillators rotate nearly independently at their natural frequency $\omega_i$. However, as the coupling becomes sufficiently strong (i.e., $\sigma$ reaches a critical threshold), all oscillators synchronize to a common frequency. To reflect the common occurrence of reciprocal interactions in oscillator applications, we employed the undirected Erd\H{o}s--R\'enyi (ER) random network as the intrinsic coupling between oscillators\upcite{RN261}. The network size $N$ ranges from 4 to 50, with each pair of nodes being independently interconnected with a probability of 0.5. The natural frequency $\omega_i$ of each oscillator is randomly selected from the interval $[0,0.05\pi]$ before shifting all natural frequencies by the same amount to transform into a corotating frame where $\sum_{i=1}^{N}\omega_i=0$. Additionally, the initial phase $\theta_i$ of each oscillator is also randomly selected from the interval $[0,0.05\pi]$. Based on these coupled equations under the above settings, we generated 100 time series by increasing the coupling strength $\sigma$ linearly from 0 to 1, with each series corresponding to a network of a different size (an example for $N=11$ is shown in Fig. 2d). Each time series includes 1000 data points, with a sampling interval of 0.01, and the first 300 data points used for training. By introducing the obtained time series into the basis functions constructed from the $\mathbf{F}$-sub-basis $\mathcal{B}_F=\{\theta_i,\nu\}$ and the $\mathbf{G}$-sub-basis $\mathcal{B}_G=\{\theta_i,\sin(\theta_j-\theta_i),\nu\}$ (degree $k=2$) where $\nu$ is set as a monotonically varying variable (i.e., $s_i\equiv +1$), we inferred governing equations for the coupled oscillators system by performing a global regression over all nodes based on the optimal basis functions. For the case in shown Fig. 2d, we inferred the true equations based on the optimal set of basis functions identified through the grid search (Fig. 2e), and for all 100 time series, our method achieved accurate dynamics prediction using the inferred equations (Fig. 2f). These results demonstrate that our method accurately recovers the equations of the Kuramoto model from data.

\begin{figure}[htbp!]
\centering
\includegraphics[width=\textwidth]{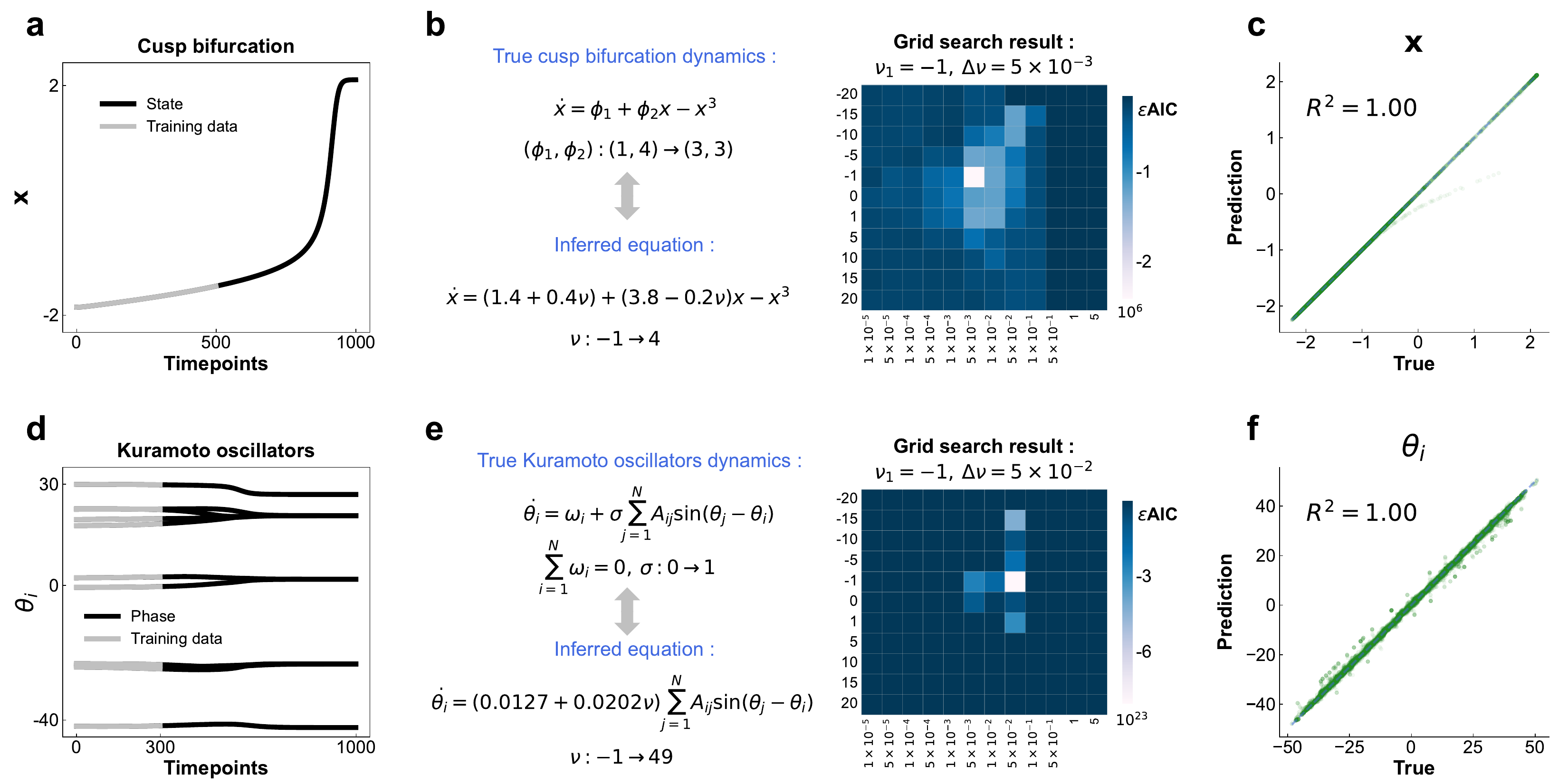}
\caption*{\fontsize{10}{12}\selectfont \textbf{Fig.2 \textbar\ Inferring dynamics of two synthetic systems.} (\textbf{a}). An example time series of the cusp bifurcation with the initial parameter values $(\phi_1,\phi_2)=(1,4)$. (\textbf{b}). The inferred governing equation for the example shown in (\textbf{a}), based on the optimal basis functions. (\textbf{c}). Comparison between the predicted and true states of the cusp bifurcation, illustrating results from 100 time series. (\textbf{d}). An example time series of the coupled Kuramoto oscillators with a network of size $N=11$. (\textbf{e}). The inferred coupled equations for the example shown in (\textbf{d}), based on the optimal basis functions. (\textbf{f}). Comparison between the predicted and true phases of the oscillators in the Kuramoto model, illustrating results from 100 time series.}
\end{figure}

\subsection{Robustness tests on large or deficient basis function library}
Intuitively, for systems whose dynamics are governed by highly complex underlying equations, a large-scale basis function library is required in order to get convincing results for inference. However, due to the fact that an excessively large basis function set often increases the risk of overfitting, the performance of inference is hampered when the cardinality of the basis function set is too large. To verify the robustness of our method against increasingly large basis function libraries, we applied it to discover governing equations over extensive libraries with various sizes. The inference inaccuracy is quantified by symmetric mean absolute percentage error (sMAPE), which measures the errors in both the function terms and coefficients between the inferred equations and the ground truth. The evaluation was conducted on the time series generated from the cusp bifurcation and the coupled Kuramoto oscillators. For the cusp bifurcation, we constructed four sets of basis functions from the sub-basis $\mathcal{B}=\{\bm{x},\nu\}$, with each set consisting of all monomials from degree 0 to 3, 4, 5, and 6, respectively. We then tested our method using these four sets of basis functions on time series generated by the cusp bifurcation equation, and compared the results with those obtained from basis functions constructed from the sub-basis $\mathcal{B}=\{\bm{x},\bm{\Phi}\}$ and the sub-basis $\mathcal{B}=\{\bm{x},t\}$. Moreover, we conducted experiments with varying training lengths to increase the diversity of experimental settings. Fig. 3a illustrates the variation in inference inaccuracy of the equations inferred through different driving factors as the basis function library expands, using 500, 550, 600, 650, and 700 data points, respectively, before the system's cusp bifurcation as training data. The results demonstrate that equations inferred using the optimal driving variable are more reliable than those inferred using the forcing parameters and the time variable, even though the inference accuracy decreases with increasing basis function complexity. For the coupled Kuramoto oscillators, we constructed four sets of basis functions from the $\mathbf{F}$-sub-basis $\mathcal{B}_F=\{\theta_i,\nu\}$ and the $\mathbf{G}$-sub-basis $\mathcal{B}_G=\{\theta_i,\sin(\theta_j-\theta_i),\nu\}$, with each set consisting of all monomials from degree 0 to 2, 3, 4, and 5, respectively. We employed 300, 350, 400, 450, and 500 data points as the training data, respectively, to infer the coupled equations, and compared the inference inaccuracy with that obtained from the basis functions with the forcing parameter and the time variable. As shown in Fig. 3b, our method exhibits significantly higher accuracy in equation inference than the methods based on either the forcing parameter or the time variable across all sizes of the basis function library, maintaining its robustness as the library expands. This further suggests the inherent advantage of the optimal driving variable, which induces a set of basis functions that enables the inferred equation to effectively capture the dynamics within the model space.

Even with extensive enlargement of the basis function library, a finite library may remain insufficient to cover all the functional components of the unknown governing equations. To evaluate the robustness of our method under a deficient basis function library, we deliberately removed key dynamical terms from the library and conducted experiments on two synthetic systems. The inferred surrogate equations were then used to predict the dynamics of the study system for evaluating their fidelity in capturing dynamics. The prediction inaccuracy is measured by the normalized Euclidean distance (NED), which quantifies the discrepancy between predicted and true trajectories. We constructed the basis functions from the sub-basis $\mathcal{B}=\{\bm{x},\nu\}$ (degree $k=2$) for the cusp bifurcation, where the cubic term has been removed. For the 100 test time series, all the surrogate equations inferred using the optimal driving variable exhibit bifurcations (Fig. 3c). Moreover, these inferred surrogate equations closely match the ground truth in both the transition location and the pre-bifurcation trajectories (Figs. 3d and 3e). In contrast, the surrogate equations, inferred using the forcing parameters and the time variable, exhibit bifurcations in 19\% and 70\% of the 100 cases, respectively. Besides, these surrogate equations that exhibit bifurcations fail to accurately capture either the location of the bifurcation points or the pre-bifurcation trajectories. For the Kuramoto oscillators, we replaced the interaction term $\sin(\theta_j-\theta_i)$ with $\sin(\theta_j+\theta_i)$ or $\cos(\theta_j-\theta_i)$ in $\mathbf{G}$-sub-basis when constructing the basis functions. Nevertheless, the predicted trajectories generated by the surrogate equations inferred using the optimal driving variable maintain a low error relative to the ground truth, and significantly outperform those based on the forcing parameters or the time variable, as shown in Figs. 3f and 3g. These results suggest that our method can accurately capture the dynamics even when using a deficient basis function library, which remains useful for revealing the underlying mechanisms of real-world systems.

\begin{figure}[htbp!]
\centering
\includegraphics[width=\textwidth]{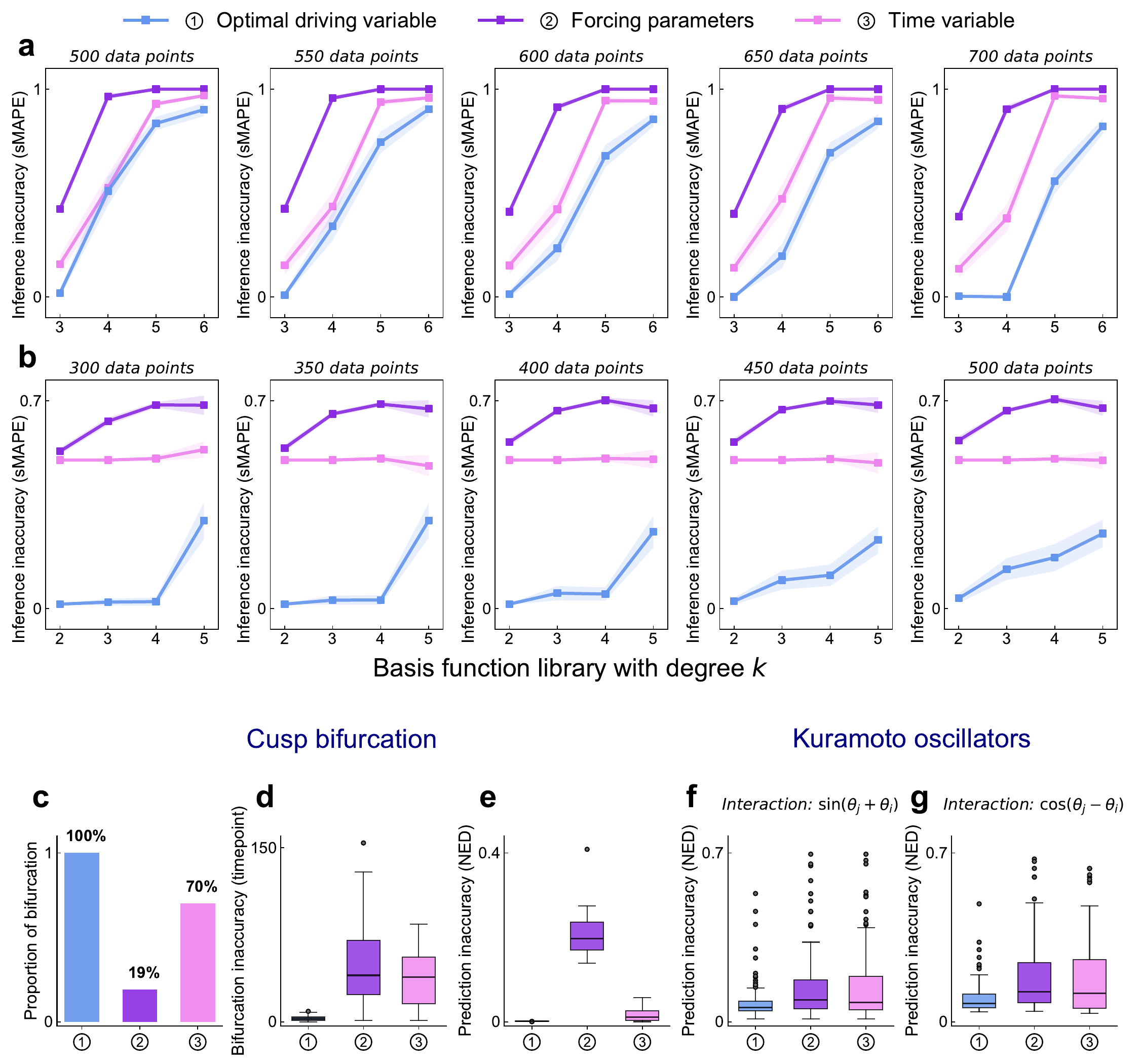}
\caption*{\fontsize{10}{12}\selectfont \textbf{Fig.3 \textbar\ Robustness tests.} (\textbf{a},\textbf{b}). Inference inaccuracy of the inferred equations for cusp bifurcation (\textbf{a}) and coupled Kuramoto oscillators (\textbf{b}), respectively, using four basis function libraries with increasingly large sizes. (\textbf{c}-\textbf{e}). Performance of dynamics inference for the cusp bifurcation under a deficient basis function library, assessed by the bifurcation detection rate (\textbf{c}), the accuracy of bifurcation point localization (\textbf{d}), and the prediction accuracy for the pre-bifurcation trajectories (\textbf{e}).}
\end{figure}

\subsection{Inferring dynamics of empirical systems}
Real-world systems often lack effective empirical models for understanding their behaviors, especially when systems are non-autonomous with highly complex dynamics. In this study, we applied our method to a wide range of real-world systems that lack expert knowledge about their underlying dynamics. These empirical systems range from cellular processes to population migration, and encompass both natural and artificial systems: (i) the ATP concentration in the cytosol of living leaf cells under hypoxia\upcite{RN249}, (ii) the UAV autonomous flight trajectory measured from our experiment, (iii) the spontaneously beating chick-heart aggregates following treatment with a potassium channel blocker\upcite{RN108}, and (iv) the seasonal abundance of 14 dominant fish species and their interactions in Maizuru Bay\upcite{RN204}. We distill the dynamical equations of these systems from their observed data, aiming to uncover the underlying governing laws. Since the governing equations of these real-world systems are unknown, we validated the accuracy of our findings by using the identified equations to predict the system dynamics and comparing the predictions with the ground truth. In addition, since the dynamics of empirical systems may not be sparsely representable in the coordinate system in which they are measured\upcite{RN256}, we not only used LASSO regression to identify their sparse dynamics, but also employed ridge regression within our optimal basis inference framework to infer the equations for these systems.

\subsubsection{Cellular energy depletion}
To investigate cellular energy dynamics, we analyzed the ATP concentration in the cytosol of \emph{Arabidopsis thaliana} leaves, as measured by S.Wagner et al.\upcite{RN249} using a genetically encoded fluorescent protein-based sensor\upcite{RN264} in a sealed 96-well multititer plate filled with medium. The ATP concentration is quantified by the yellow-to-blue fluorescence emission ratio of the sensor (referred to as the relative fluorescence ratio). The experiment was performed under dark conditions to eliminate oxygen production from photosynthesis. Consequently, the leaf tissue respired off the dissolved oxygen, gradually establishing hypoxia in the medium. After initiating fluorescence recording, the sensor indicated a slight gradual decline in cytosolic ATP up to a tipping point, at which the cytosolic ATP concentration suddenly decreased. The experiment, illustrated in Fig. 4a, resulted in a dataset of 271 data points spanning a total of 3 minutes, with a time interval of 0.011 minutes. We utilized the time series data of the first 2 minutes for training, and then tested our method based on the basis functions constructed from the sub-basis $\mathcal{B}=\{\bm{x},\nu\}$ (degree $k=3$). Here we defined $\nu$ as a monotonically varying variable (i.e., $s_i\equiv+1$) to represent the gradual depletion of oxygen in the medium. Based on the inferred dynamical equation, we successfully predict the collapse of the ATP concentration (Fig. 4b). Moreover, we used AUTO-07P\upcite{RN93} to analyze this inferred dynamical equation, identifying that the equation exhibits a fold bifurcation at $\nu=1.24$ (Time $=4.69$~min). Our results provide the dynamical mechanism underlying the abrupt drop in the cytosolic ATP under hypoxia, where a fold bifurcation occurs.

\subsubsection{UAV autonomous flight}
In mobile robotics, autonomous navigation is the process by which a robot moves from its starting position to a desired goal by continuously generating collision-free trajectories. This process operates in a closed-loop manner: the robot first senses the environment using onboard sensors and converts raw sensor data into a usable representation, then plans the path based on the collected information, and finally executes the decisions. Since the perceived environment changes over time, this perception--planning--control loop can be considered as a nonlinear non-autonomous system with respect to the robot's position. To investigate whether our approach can capture the dynamics underlying the flight governed by artificial control algorithms, we conducted simulation experiments using a quadcopter UAV. We collected the robot's positions during a collision-avoidance navigation task as the test dataset. Specifically, the UAV started at position $\text{P}_{\text{s}}$ and was directed toward a user-specified goal position $\text{P}_{\text{g}}$ ($\text{P}_{\text{s}}$ and $\text{P}_{\text{g}}$ lie on the same horizontal plane). Along the flight path from $\text{P}_{\text{s}}$ to $\text{P}_{\text{g}}$, two obstacles were positioned alternately on either side of the path, and the UAV was expected to reach $\text{P}_{\text{g}}$ without any collision with the obstacles, as illustrated in Fig. 4c. The UAV was equipped with an onboard depth camera capable of estimating depth information within a 5~m range. During the flight, the depth images were converted into an occupancy map\upcite{RN286}, which was then used for B-spline trajectory optimization for planning\upcite{RN285}. Meanwhile, we recorded the UAV's X and Y positions at a rate of 30~Hz, resulting in a historical trajectory of 276 position data points. We utilized the data collected by the UAV near the first obstacle as the training set to infer the non-autonomous dynamical equations based on the basis functions constructed from the sub-basis $\mathcal{B}=\{\bm{x},\nu\}$ (degree $k=3$). To capture the accumulation of overall environmental perception data in the onboard computer, $\nu$ was configured to vary monotonically (i.e., $s_i\equiv+1$). Based on the inferred equations, we predicted the UAV's flight path near the second obstacle toward the goal position $\text{P}_{\text{g}}$. The comparison between the predicted and true trajectories in Fig. 4d demonstrates that the inferred equations accurately capture the UAV's flight dynamics. Moreover, we calculated the acceleration of UAV based on our inferred equations, thereby analyzing the variation of its horizontal thrust during flight (illustrated as arrows in Fig. 4d). Specifically, the variation of the thrust acting on the UAV can be divided into four distinct phases: (i) initial phase (2.35~s to 4.89~s), where the UAV identifies potential obstacles and performs obstacle avoidance planning; (ii) the first obstacle avoidance phase (4.89~s to 9.22~s); (iii) the second obstacle avoidance phase (9.22~s to 12.64~s); and (iv) the deceleration phase (12.64~s to 12.80~s), where the UAV decelerates before reaching the goal position $\text{P}_{\text{g}}$. It is noteworthy that the thrust experienced by the UAV during phases (iii) and (iv) was entirely predicted based on our inferred equations. The predicted thrust variations align well with the actual flight scenario, and the emergence of reverse thrust for deceleration accurately predicted the user-specified goal position $\text{P}_{\text{g}}$. This indicates that our method can not only accurately predict the flight trajectory but also distill the UAV's motion control algorithms in the form of dynamical equations.

\subsubsection{Arrhythmia of beating chick-heart aggregates}
To further demonstrate the versatility of our method on discrete-time real-world systems, we inferred the equations to reveal the rhythm dynamics of spontaneously beating chick-heart aggregates following treatment with a potassium channel blocker\upcite{RN108}. The aggregates were treated with 0.5-2.5~$\mu$mol of E-4031, a compound that blocks the human Ether-\`{a}-go-go-Related Gene (hERG) potassium channel\upcite{RN298} and thereby induces arrhythmia, as illustrated in Fig. 4e. The beating of the aggregates was recorded using phase-contrast imaging and the inter-beat intervals (IBI) were computed as the time between consecutive beats. As the E-4031 diffused within the aggregates, the IBI evolved from a relatively constant pattern to a periodic pattern in which the intervals alternated between distinct values, indicating the onset of arrhythmia. The timing of this onset in chick-heart aggregates varied with the E-4031 dose. To model the dynamical process under different doses of E-4031, we inferred the dynamical equations of aggregates treated with three distinct doses, with each group yielding 220, 230, and 335 IBI recordings, respectively. The first 150 data points for the 220- and 230-recording groups, and the first 200 data points for the 335-recording group were used for training. We then tested our method based on the basis functions constructed from the sub-basis $\mathcal{B}=\{\bm{x},\nu\}$ (degree $k=3$), where $\nu$ was set as a monotonically varying variable (i.e., $s_i\equiv+1$) to reflect the diffusion of the drug within the aggregates\upcite{RN308}. Our method successfully predicts the transition in IBI patterns, from a constant rhythm to a period-2 rhythm (Figs. 4f-h). Moreover, we used AUTO-07P\upcite{RN93} to analyze the three inferred dynamical equations, which identifies that each equation exhibits a period-doubling bifurcation at $\nu=$ 0.973, -0.022, and -0.705 (corresponding to Beat number $=$ 195, 196, and 295), respectively. Our results reveal the underlying mechanism of the arrhythmia in beating chick-heart aggregates, which corresponds to a period-doubling bifurcation.

\begin{figure}[htbp!]
\centering
\includegraphics[width=1\textwidth]{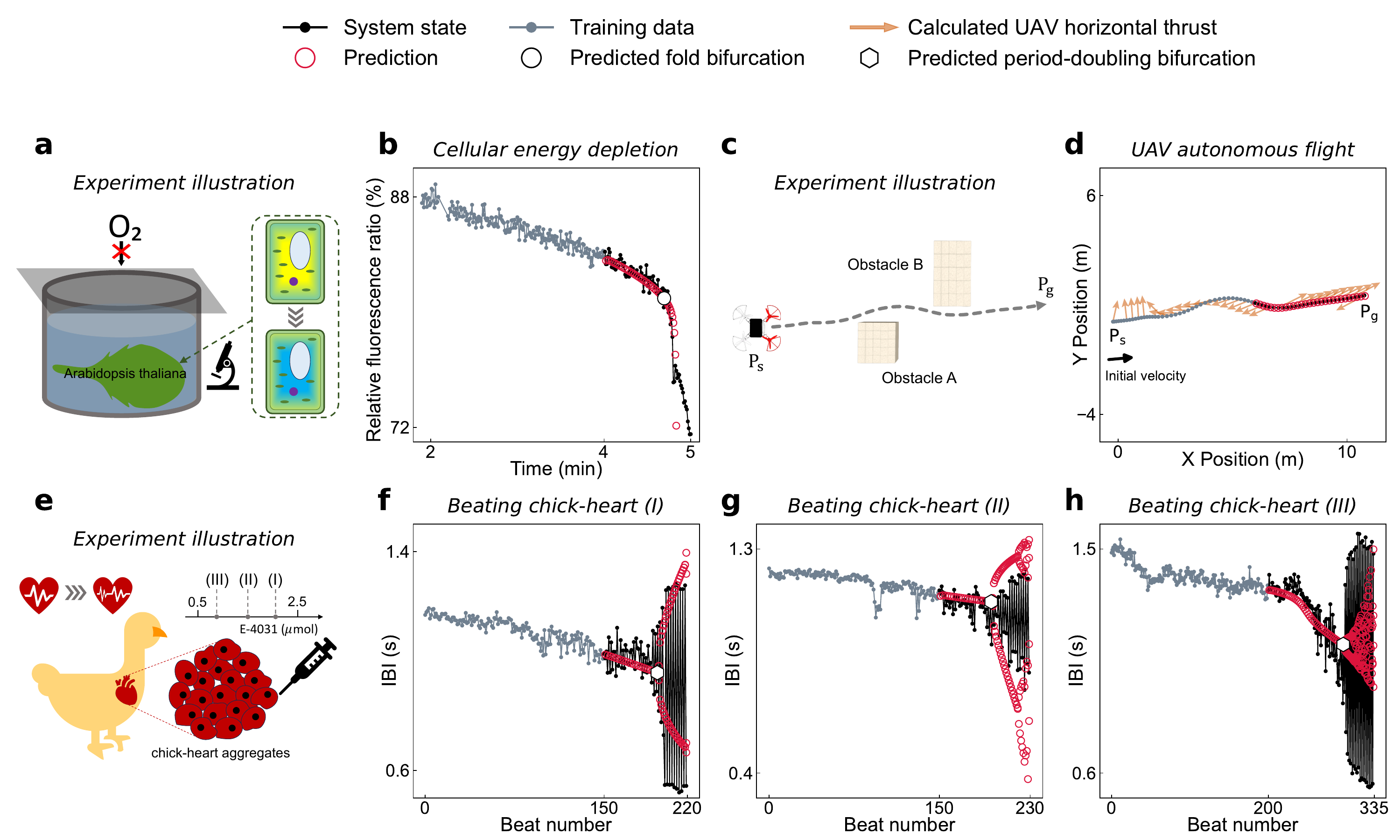}
\caption*{\fontsize{10}{12}\selectfont \textbf{Fig.4 \textbar\ Predicting dynamics of empirical systems derived from cellular biology, control engineering, and cardiac physiology.} (\textbf{a}). Energy depletion in the cytoplasm of leaf cells under dark and hypoxic conditions\upcite{RN249}; ATP concentration calculated from the relative fluorescence ratio. The decrease in cellular ATP concentration is indicated by a lower yellow-to-blue fluorescence emission ratio, resulting in a fluorescence shift from yellow to blue. (\textbf{b}). Comparison between the predicted and true trajectories of ATP concentration. The predicted trajectory is generated by the inferred dynamical equation, in which a fold bifurcation is identified. (\textbf{c}). Illustration of the UAV navigation experiment. (\textbf{d}). Comparison between the predicted and true UAV trajectories and the thrust variations during autonomous flight. (\textbf{e}). Arrhythmia of chick-heart aggregates under three different doses of E-4031\upcite{RN108}. (\textbf{f}-\textbf{h}). Comparison between the predicted and true trajectories of inter-beat intervals (IBI) for consecutive beats of the chick-heart aggregates treated with three different doses of E-4031. Period-doubling bifurcations are identified in these three inferred dynamical equations.}
\end{figure}

\subsubsection{Marine fish community}
The ability to understand and predict the dynamics of complex ecological systems is critical to species conservation amidst escalating global environmental change\upcite{RN304}. As the final case, we uncovered the population dynamics of a natural marine fish community in Maizuru Bay, Japan\upcite{RN204}. This dataset consists of the temporal abundance of 14 dominant fish species across 285 fortnightly observations spanning approximately 12 years, along with their directed interaction network (see Fig. 5a). The marine fish migration and growth are primarily influenced by two external factors: water temperature and salinity\upcite{RN253,RN265}. Since the salinity does not exhibit clear seasonality in this bay\upcite{RN204}, the water temperature can be regarded as the forcing parameter for this non-autonomous system. The data collected between 2002 and 2012 served as training samples to predict the changes in population size over the subsequent two years. Based on commonly accepted population dynamics models\upcite{RN86,RN268,RN269}, we constructed the basis functions from the $\mathbf{F}$-sub-basis $\mathcal{B}_F=\{\bm{x}_i,\frac{1}{1+\bm{x}_i},\frac{1}{1+e^{-\bm{x}_i}},\nu\}$ and the $\mathbf{G}$-sub-basis $\mathcal{B}_G=\{\bm{x}_i,\bm{x}_j,\frac{1}{1+\bm{x}_i},\frac{1}{1+\bm{x}_j},\frac{1}{1+e^{-\bm{x}_i}},\frac{1}{1+e^{-(\bm{x}_j-\bm{x}_i)}},\nu\}$ (degree $k=3$) to infer the underlying equations for this fish community. Since the population dynamics are heterogeneous\upcite{RN310,RN311,RN312}, suggesting that different species have distinct self dynamics and interspecies interactions, we inferred the dynamics for each fish species separately. To synchronize the variation of the driving variable $\nu$ with the annual periodic fluctuation in water temperature, we modulated its value to track the temperature variation. Specifically, we set $\nu=\nu_1$ in early February (the coldest period of the year in this bay) and increased $\Delta \nu$ every fortnight (i.e., $s_i=+1$) until early August (the hottest time of the year in this bay), after which we decreased $\Delta \nu$ every fortnight (i.e., $s_i=-1$) until early February. Here, the values of $\nu_1$ and $\Delta \nu$ for the optimal driving variable are determined through a grid search. We predicted the population size of each species through the inferred population dynamical equations. The prediction results for each fish species, as illustrated in Fig. 5b, show remarkable consistency with the real data. Notably, the model accurately captures the seasonal patterns underlying population dynamics, even for the few species with relatively larger prediction errors. This suggests that the equations inferred by our method reveal the underlying evolutionary mechanisms of this marine fish community. To further substantiate the fidelity of our inferred equations, we first computed the variance of the total populations across all fish species using a rolling window of length 10 (after normalization), which represents the temporal population fluctuations observed in the fish community. Subsequently, we assessed the temporal dynamic stability of this discrete-time system by computing the modulus of the dominant eigenvalue of the Jacobian matrix derived from the inferred equations at each time point, where a higher modulus indicates lower dynamic stability. These two metrics are shown to be significantly positively correlated by Kendall's rank correlation test\upcite{RN267} ($p\text{-value}=7.49\times10^{-5}$), indicating that the inferred equations accurately capture the fluctuation pattern of the system, thereby supporting their validity and reliability. Moreover, the seasonal patterns in dynamic stability observed in this fish community broadly support expectations of current ecological theory\upcite{RN204}.

\begin{figure}[htbp!]
\centering
\includegraphics[width=\textwidth]{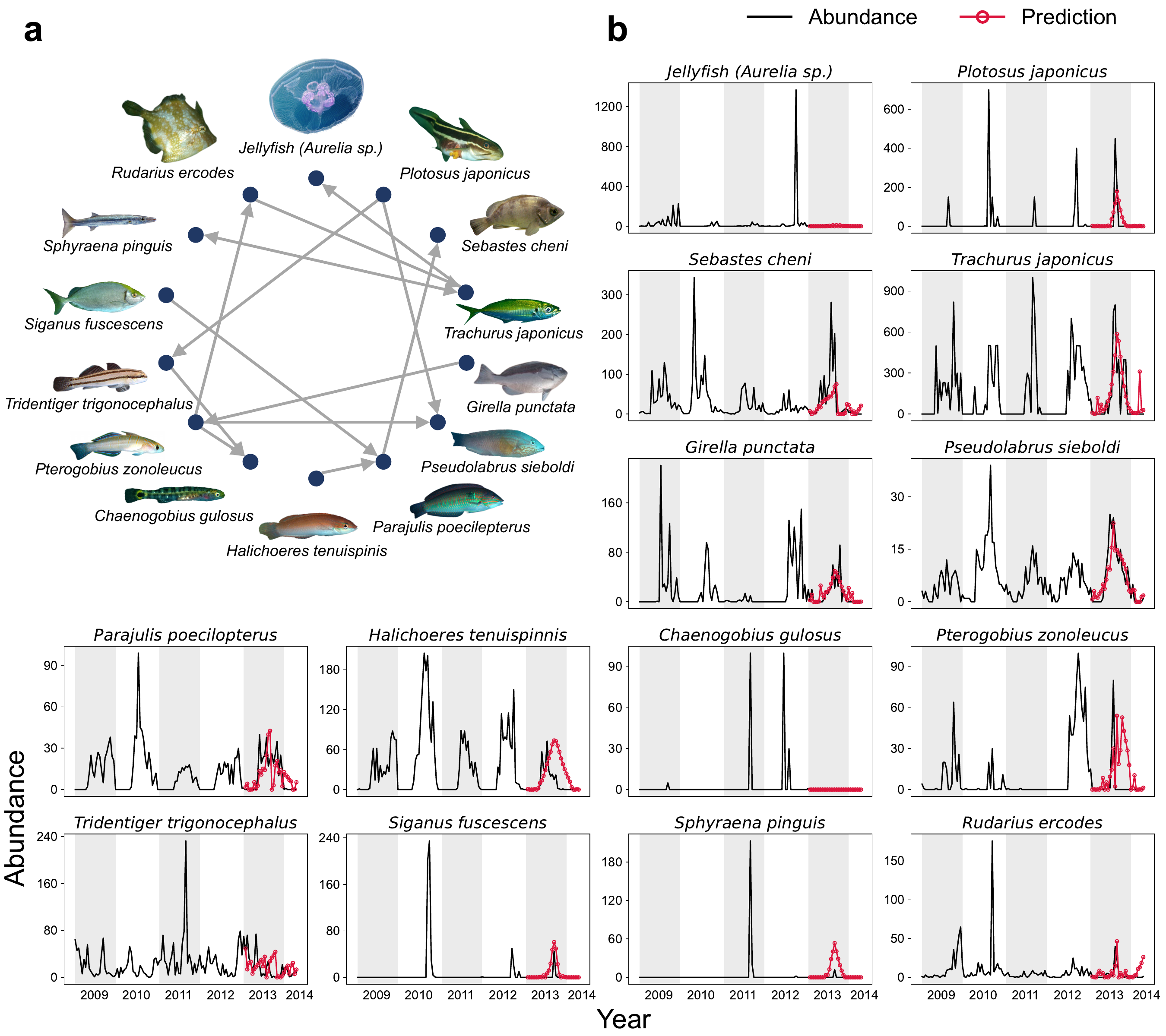}
\caption*{\fontsize{10}{12}\selectfont \textbf{Fig.5 \textbar\ Predicting the population dynamics of the marine fish community.} (\textbf{a}). Marine fish community of 14 dominant species interacting through a directed network\upcite{RN204}. (\textbf{b}). Comparison between the predicted and the true trajectory, where the predicted trajectory is generated by the inferred dynamical equations. For better visualization, the time series of population abundance are displayed during 2009 and 2014. The width of the grey region corresponds to a 1-year interval (24 observations per year).}
\end{figure}

\subsection{Comparative studies}
To demonstrate the superiority in accuracy when inferring dynamics using the optimal set of basis functions within the model space, we compared the inference accuracy of the two synthetic systems and four empirical systems with those obtained from basis functions constructed using the forcing parameters and the time variable. For the synthetic systems with known governing equations, we evaluated the inference accuracy at each time point by comparing the inferred equations against the ground truth. The results from both the cusp bifurcation (Fig. 6a) and coupled Kuramoto oscillators (Fig. 6b) consistently demonstrate that our method significantly outperforms methods based on the forcing parameters and the time variable. Meanwhile, as for the cusp bifurcation, we noted that the accuracy of equations inferred based on multiple forcing parameters is significantly lower. This disparity in accuracy stems from the inference difficulty caused by the redundant terms in the basis function library under multiple parameters, while our method circumvents this issue by constructing a complete basis for the model space. To evaluate the fidelity of the inferred equations for real-world systems with unknown governing dynamics, we calculated the normalized Euclidean distance (NED) between the predicted trajectories generated by these equations and the true trajectories. For the empirical systems with unobservable forcing parameters, including cellular energy, UAV navigation, and chick-heart aggregates, we compared the prediction accuracy of the equations inferred using the optimal basis functions with that of the equations inferred with the time variable. For the marine fish community, we conducted comparative tests between our method and those based on the time variable and the accessible forcing parameter (water temperature). The comparative results presented in Figs. 6c-h further confirm the superior accuracy of dynamics inference using the optimal basis functions. This superiority is manifested in both the smaller prediction errors achieved by our method and its ability to successfully capture bifurcations.

In identifying the optimal basis functions for dynamics inference, our framework incorporates a key component, the optimization of numerical errors. This numerical error originates from the matrix pseudo-inverse calculation used in solving the coefficient matrix. To study how the numerical errors influence the inferred results, we conducted an ablation study in which the numerical error optimization was disabled during the dynamics inference process. Specifically, we compared the results obtained using the optimal basis functions with the least $\varepsilon$AIC to those with the least original AIC. As shown in Fig. 6, the results obtained by the equations inferred via the basis functions based on AIC (cyan-colored curves) are significantly less accurate than those based on $\varepsilon$AIC. This finding highlights the critical role of numerical error optimization in achieving reliable dynamics inference, a factor that has been overlooked in previous studies. Our framework explicitly incorporates numerical error optimization, thereby establishing a more reliable foundation for the discovery of dynamical equations.

\begin{figure}[htbp!]
\centering
\includegraphics[width=\textwidth]{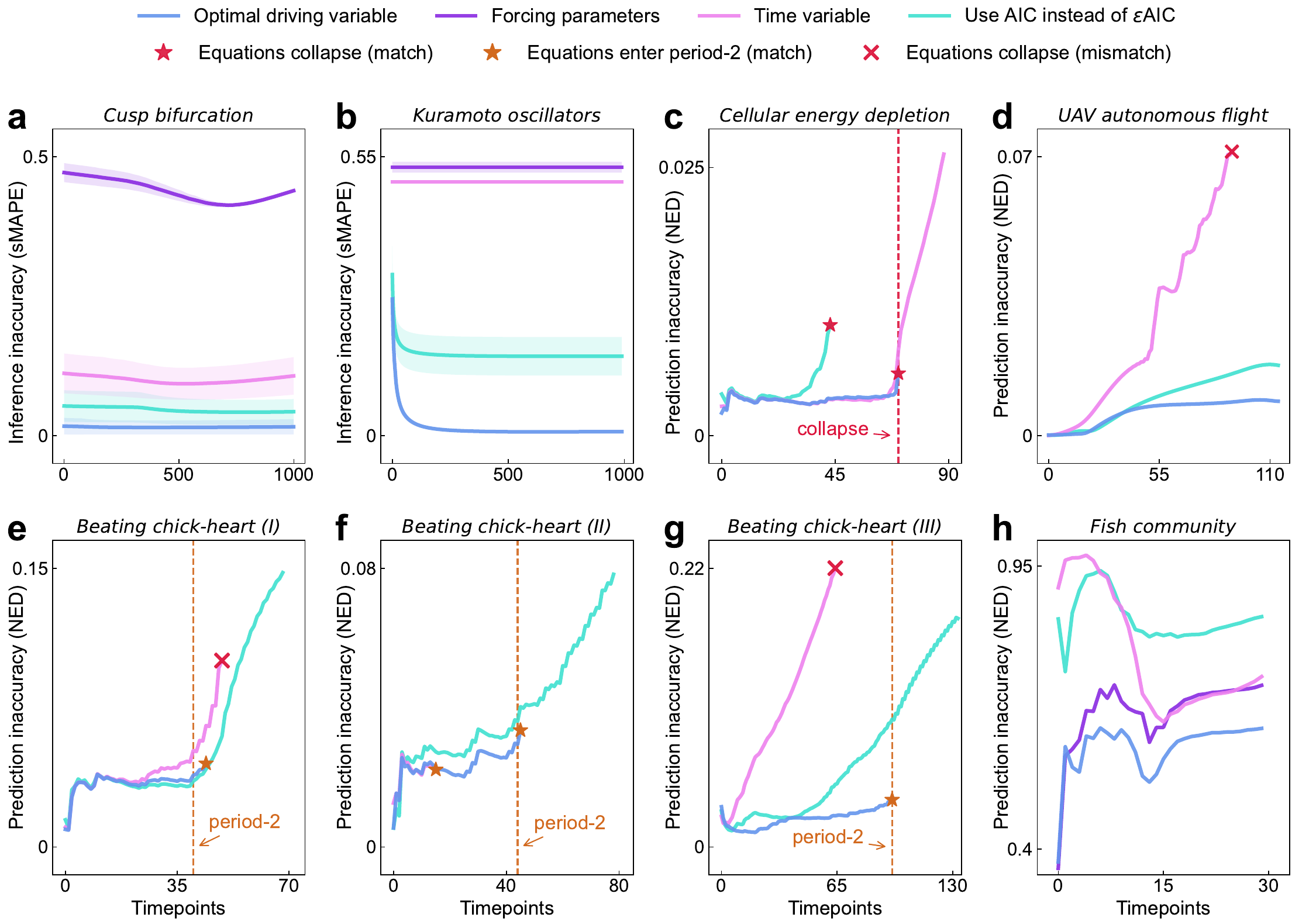}
\caption*{\fontsize{10}{12}\selectfont \textbf{Fig.6 \textbar\ Comparative studies.} Two comparative studies are performed: the comparison of the basis functions based on the optimal driving variable with those based on the forcing parameters and the time variable, and the comparison of the optimal basis functions determined by the least $\varepsilon$AIC with those determined by the least original AIC. (\textbf{a},\textbf{b}). Inference inaccuracy of the cusp bifurcation (\textbf{a}) and the coupled Kuramoto oscillators (\textbf{b}). (\textbf{c}-\textbf{h}). Prediction inaccuracy of real-world systems are illustrated for the cellular energy status under hypoxia (\textbf{c}), the UAV autonomous flight trajectory (\textbf{d}), the chick-heart aggregates treated with three different doses of E-4031 (\textbf{e}-\textbf{g}), and the natural marine fish community (\textbf{h}).}
\end{figure}

\section{Discussion}
In this work, we prove the equivalence theorem for the basis functions, and based on this, we propose an effective data-driven method to reveal the dynamics of non-autonomous complex systems by inferring their equations. The effectiveness of this method lies in identifying a set of basis functions with an optimal driving variable from a given model space, enabling the inference of equations that best characterize the system's underlying dynamics. Our approach not only simplifies multiple forcing parameters to a single driving variable, but more importantly, balances accuracy (including both numerical and fitting accuracy) and model complexity in equation inference. The experimental results demonstrate the effectiveness of our approach in equation discovery on synthetic systems and highlight its ability to uncover the complex dynamics of real-world systems using simple basis function libraries.

Our method markedly broadens real-world applicability of dynamics inference, since it is capable of uncovering complex non-autonomous dynamics under simplified prior specifications. This simplification involves two main aspects: (i) the observed time series of forcing parameters is not required for dynamics inference; (ii) only minimal prior knowledge about the true dynamics is needed to infer them through a simple basis function library. For instance, when inferring real-world biochemical or flight dynamics, there is no requirement of any observed data of the forcing parameters, while these parameters latently drive system's evolution. In this case, our approach faithfully reconstructs the underlying dynamics and accurately captures bifurcations merely based on monomial basis functions. For nearly all complex systems in the real world, we lack comprehensive expert knowledge or observable data on their forcing parameters. In response, our proposed approach provides a solution to this universal challenge.

Even in the complete absence of prior information about the forcing parameters (including their trends), effective dynamics reconstruction can still be achieved within our framework by casting the unknown parameters as nonlinear functions of time variable. Nonetheless, we recommend having a qualitative understanding of the trends in the forcing parameters (i.e., the sign sequence $\{s_i\}_{i=1}^{T-1}$). As shown by the empirical examples in this work, this requirement is well motivated for real-world complex systems, where forcing parameters often exhibit regular temporal variation, either as monotonic changes governed by the underlying physical or biological processes, or as periodic variations that follow seasonal patterns. Based on our experience, a qualitative grasp of the forcing parameter trends can enable more robust recovery of non-autonomous dynamics. For real-world complex systems, such as human mood systems\upcite{RN315} and financial markets\upcite{RN314}, in which the forcing parameters are often elusive, how to infer changes in the forcing parameters and the resulting non-autonomous dynamics from state time series remains an open question.

For data-driven inference of governing equations, prior work has predominantly focused on estimating the optimal coefficients for a predefined set of basis functions within the model space. Recent state-of-the-art methods, such as LaGNA\upcite{RN246} and BILLIE\upcite{RN292}, have leveraged advanced machine learning techniques to extensively explore the potential of this paradigm for uncovering underlying dynamics. In contrast, our work diverges from this paradigm by identifying a driving variable to determine an optimal set of basis functions within the model space. Extensive experimental results demonstrate that our framework provides a versatile and effective approach for inferring the governing equations of non-autonomous systems. Beyond estimating optimal coefficients, our approach adaptively identifies the optimal basis functions, opening a new avenue for inferring the dynamics of complex systems. The potential of this novel methodological perspective for dynamics inference warrants further exploration in future studies. For example, since the linear dynamics on hypergraphs can always be rewritten as a dynamics on a surrogate pairwise network\upcite{RN321}, an interesting application of our paradigm is to adaptively identify the optimal basis functions under a surrogate network to infer the linear dynamics.

\section{Methods}
\subsection{Construction of basis functions for non-autonomous dynamics}
To construct basis functions, we define a sub-basis $\mathcal{B}\coloneqq \{g_1(\bm{x}),\dots,g_z(\bm{x}),\bm{u}\}$ where $g_i$ are functions of only the state variables $\bm{x}=(x_1,\dots,x_d)$ and $\bm{u} \in \{\bm{\Phi},\nu\}$ is the driving factor. Functions $g_i(\bm{x})\coloneqq (g_i(x_1),\dots,g_i(x_d))$ in the sub-basis are chosen as elementary building blocks, such as $\bm{x}$ and $\sin(\bm{x})$, rather than composite functions formed through products (e.g. $\bm{x}^2$, $\bm{x}\sin(\bm{x})$). This construction avoids redundancy in the function library, since all basis functions are linearly independent. Here, the basis functions consist of all monomials from degree 0 to $k$, constructed from the elements of the sub-basis. For example, the set of basis functions
\begin{equation*}
\begin{aligned}
\{&1,g(x_1),g(x_2),\phi_1,\phi_2,g^2(x_1),\phi_1g(x_1),\phi_2g(x_1),\\
&g^2(x_2),\phi_1g(x_2),\phi_2g(x_2),g(x_1)g(x_2),\phi_1^2,\phi_1\phi_2,\phi_2^2\}
\end{aligned}
\end{equation*}
are constructed from the sub-basis $\mathcal{B}=\{g(\bm{x}),\bm{\Phi}\}$ (degree $k=2$), where $\bm{x}=(x_1,x_2)$ and $\bm{\Phi}=(\phi_1,\phi_2)$. We constrain the driving factor to appear exclusively as monomial terms within the basis functions, instead of the heuristic approach of introducing the driving factor into the basis function library in a variety of non-monomial forms. This constraint reduces the complexity of the library while maintaining the ability to represent complex dynamics\upcite{RN216,RN294}, since any function can be expanded in a power series with respect to the driving factor.

For the system with network topology $A_{ij}$ and $N$ nodes, the dynamics are captured by coupled equations\upcite{RN262,RN255,RN309}
\begin{equation*}
\dot{\bm{x}}_i=F_i(\bm{x}_i)+\sum_{j=1}^{N}A_{ij}G_{ij}(\bm{x}_i,\bm{x}_j),
\end{equation*}
which include the self dynamics $F_i$ and interaction dynamics $G_{ij}$. We construct two separate libraries of basis functions to reconstruct the self dynamics $F_i$ and the interaction dynamics $G_{ij}$\upcite{RN243,RN246}, respectively. Here, we refer to the sub-basis that construct basis functions for $F_i$ and $G_{ij}$ as $\mathbf{F}$-sub-basis $\mathcal{B}_F$ and $\mathbf{G}$-sub-basis $\mathcal{B}_G$, respectively. Notably, network dynamics can be further categorized into homogeneous dynamics and heterogeneous dynamics. For homogeneous dynamics, for example those in coupled oscillators\upcite{RN260}, all nodes share the same self dynamics, and all pairwise interactions follow an identical rule. That is, $F_i\equiv F$ and $G_{ij}\equiv G$. Therefore, we perform a global regression over all nodes to infer the governing equations of the homogeneous dynamics. In contrast, heterogeneous dynamics characterize systems in which nodes exhibit distinct dynamical behaviors and their interactions are governed by diverse rules, such as in population dynamics\upcite{RN310,RN311,RN312}. In this case, we infer the heterogeneous dynamics by performing regression for each node separately.

\subsection{Equivalence theorem for basis functions}
First, we demonstrate that the model space spanned by the basis functions with $\bm{\Phi}=(\phi_1,\dots,\phi_n)$ is identical to that spanned by the basis functions with $\nu$. It is observed that at the $i$-th time point, the $j$-th component of $\bm{\Phi}$ takes the value $\phi_{j,i}$, and satisfies $\phi_{j,i+1}=\phi_{j,i}+s_i\Delta\phi_j$, while $\nu$ takes the value $\nu_{i}$ and satisfies $\nu_{i+1}=\nu_{i}+s_i\Delta \nu$. By extracting the common factor $s_i$ from both expressions, we have
\begin{equation}
    s_i=\frac{\phi_{j,i+1}-\phi_{j,i}}{\Delta \phi_j}=\frac{\nu_{i+1}-\nu_i}{\Delta\nu}.
    \label{eq:1}
\end{equation}
Summing Eq.~\eqref{eq:1} over $i=1,\dots,m$, we obtain, for all $m$,
\begin{equation}
    \frac{\phi_{j,m+1}-\phi_{j,1}}{\Delta \phi_j}=\frac{\nu_{m+1}-\nu_1}{\Delta\nu}.
    \label{eq:2}
\end{equation}
Eq.~\eqref{eq:2} demonstrates the linear relationship between $\phi_{j}$ and $\nu$, namely,
\begin{equation}
    \phi_j=\frac{\Delta\phi_j}{\Delta\nu}(\nu-\nu_1)+\phi_{j,1},
    \label{eq:3}
\end{equation}
where $\phi_{j,1}$ and $\Delta \phi_j$ are experimentally observed constants, while $\nu_1$ and $\Delta \nu$ are arbitrary constants. Let $\{\psi_i(\bm{x},\bm{\Phi})\}_{i=1}^P$ and $\{\psi_j(\bm{x},\nu)\}_{j=1}^Q$ denote two sets of basis functions constructed from sub-bases $\{g_1(\bm{x}),\dots,g_z(\bm{x}),\bm{\Phi}\}$ and $\{g_1(\bm{x}),\dots,g_z(\bm{x}),\nu\}$, respectively, where $P\geq Q$ and $g_1,\dots,g_z$ are $z$ linearly independent functions. Since the linear relationship Eq.~\eqref{eq:3} between $\nu$ and any component of $\bm{\Phi}$, for each $i=1,\dots,P$ there exist $a_{ij}$, $j=1,\dots,Q$, such that $\psi_i(\bm{x},\bm{\Phi})=\sum_{j=1}^{Q}a_{ij}\psi_j(\bm{x},\nu)$, which holds for any $\nu_1,\,\Delta\nu \in \mathbb{R}$. Then we have
\begin{equation}
\operatorname{span}\{\psi_i(\bm{x},\bm{\Phi})\}_{i=1}^P\subseteq \operatorname{span}\{\psi_j(\bm{x},\nu)\}_{j=1}^Q.
\label{eq:4}
\end{equation}
Now we consider upgrading the inclusion to equality, which suffices to prove that $\operatorname{rank}\{\psi_i(\bm{x},\bm{\Phi})\}_{i=1}^P\geq\operatorname{rank}\{\psi_j(\bm{x},\nu)\}_{j=1}^Q$. Since $\phi_1$ is a special case of $\nu$ with $\nu_1=\phi_{1,1}$ and $\Delta \nu = \Delta \phi_1$, the basis functions constructed from $\{g_1(\bm{x}),\dots,g_z(\bm{x}),\phi_1\}$ can be denoted as $\{\psi_j(\bm{x},\phi_1)\}_{j=1}^Q$. It is observed that $\{\psi_j(\bm{x},\phi_1)\}_{j=1}^Q\subseteq \{\psi_i(\bm{x},\bm{\Phi})\}_{i=1}^P$ and $\operatorname{rank}\{\psi_j(\bm{x},\phi_1)\}_{j=1}^Q
=Q$, thus,
\begin{equation}
\operatorname{rank}\{\psi_i(\bm{x},\bm{\Phi})\}_{i=1}^P\geq Q
=\operatorname{rank}\{\psi_j(\bm{x},\nu)\}_{j=1}^Q.
\label{eq:5}
\end{equation}
Combining Eqs.~\eqref{eq:4} and~\eqref{eq:5}, we prove that
\begin{equation}
\operatorname{span}\{\psi_i(\bm{x},\bm{\Phi})\}_{i=1}^P=\operatorname{span}\{\psi_j(\bm{x},\nu)\}_{j=1}^Q,
\label{eq:6}
\end{equation}
and
\begin{equation}
\operatorname{rank}\{\psi_i(\bm{x},\bm{\Phi})\}_{i=1}^P=Q.
\label{eq:7}
\end{equation}
Given the state time series $(\mathbf{x}_1,\dots,\mathbf{x}_T)$ and the driving factor time series $(\mathbf{u}_1,\dots,\mathbf{u}_T)$, with $\bm{u}\in\{\bm{\Phi},\nu\}$, we define the feature matrix $\Theta_{\bm{u}}$ by evaluating the basis functions $\{\psi_i(\bm{x},\bm{u})\}_{i=1}^{P\,\text{or}\,Q}$ at time points $t=1,\dots,T$. Let $\mathbf{c}_{\bm{u}}^{(r)}\coloneqq \big(\psi_r(\mathbf{x}_1,\mathbf{u}_1),\dots,\psi_r(\mathbf{x}_T,\mathbf{u}_T)\big)^{\top}$ denote the $r$-th column vector of the feature matrix $\Theta_{\bm{u}}$. Then
\begin{equation}
\operatorname{Col}\big(\Theta_{\bm{\Phi}}\big)=\operatorname{span}\{\mathbf{c}_{\bm{\Phi}}^{(i)}\}_{i=1}^P,\quad
\operatorname{Col}\big(\Theta_{\nu}\big)=\operatorname{span}\{\mathbf{c}_{\nu}^{(j)}\}_{j=1}^Q.
\label{eq:8}
\end{equation}
Combining Eqs.~\eqref{eq:6},~\eqref{eq:7} and~\eqref{eq:8}, then the following theorem holds.

\vspace{10pt}
\noindent \textbf{Theorem.} For all $\nu_1,\,\Delta \nu \in \mathbb{R}$,
\begin{equation*}
\operatorname{span}\{\psi_i(\bm{x},\bm{\Phi})\}_{i=1}^P= \operatorname{span}\{\psi_j(\bm{x},\nu)\}_{j=1}^Q,\quad \operatorname{Col}\big(\Theta_{\bm{\Phi}}\big)=\operatorname{Col}\big(\Theta_{\nu}\big).
\end{equation*}
Moreover, $\dim \big(\operatorname{span}\{\psi_j(\bm{x},\nu)\}_{j=1}^Q\big)=\dim \big(\operatorname{Col}\big(\Theta_{\nu}\big)\big)=Q$.
\vspace{10pt}

Here, we note that the set of basis functions $\{\psi_i(\bm{x},\bm{\Phi})\}_{i=1}^P$ which contains $P$ elements will be linearly dependent if $P>Q$. According to the theorem, the rank of the model space spanned by this set is $Q$, meaning that even though the set contains $P$ elements, only $Q$ of them provide independent features, and the rest are redundant. In contrast, the set of basis functions associated with $\nu$ forms a complete basis for the model space, with its cardinality equal to the rank of the model space.

Furthermore, we demonstrate that the inferred equations with these sets of basis functions, obtained from solving $\dot{\mathbf{X}}=\Theta\mathbf{A}$ via least squares, are theoretically equivalent, where $\dot{\mathbf{X}}=(\dot{\mathbf{x}}_1,\dots,\dot{\mathbf{x}}_T)^{\top}$. In terms of the geometric description of least squares estimation, $\Theta\mathbf{A}$ provides an estimate of $\dot{\mathbf{X}}$, which represents the orthogonal projections of the columns of $\dot{\mathbf{X}}$ onto the column space of $\Theta$. That is,
\begin{equation*}
    \Theta^{\top}(\dot{\mathbf{X}}-\Theta\mathbf{A})=0.
\end{equation*}
We denote the least squares solutions for coefficient matrices corresponding to $\nu$ and $\bm{\Phi}$ as $\mathbf{A}_\nu$ and $\mathbf{A}_{\bm{\Phi}}$, respectively. Based on this theorem, the solved governing equations for $\nu$ and $\bm{\Phi}$ are identical, given by
\begin{equation*}
\dot{\bm{x}}=\Psi_{\nu}\mathbf{A}_\nu=\Psi_{\bm{\Phi}}\mathbf{A}_{\bm{\Phi}},
\end{equation*}
where $\Psi_{\nu}=(\psi_1(\bm{x},\nu),\dots,\psi_Q(\bm{x},\nu))$ and $\Psi_{\bm{\Phi}}=(\psi_1(\bm{x},\bm{\Phi}),\dots,\psi_P(\bm{x},\bm{\Phi}))$. A similar equivalence holds when using the least absolute shrinkage and selection operator (LASSO). Since LASSO can be replaced by the sequential thresholded least squares method\upcite{RN216}, which often yields the same sparsity pattern.

\subsection{Numerical approximation in least squares estimation}
The regularized least-squares estimate of $\mathbf{A}$ is given by
\begin{equation*}
    \mathbf{A}=(\Theta^{\top}\Theta+\alpha\mathbf{I})^{-1}\Theta^{\top}\dot{\mathbf{X}},
\end{equation*}
where $\alpha$ is the regularization parameter, $\alpha=0$ corresponds to the ordinary least-squares case, and $\mathbf{I}$ is the identity matrix. In practice, $\mathbf{A}$ is computed numerically in the pseudo-inverse form $\mathbf{A}=(\Theta^{\top}\Theta+\alpha\mathbf{I})^{\dagger}\Theta^{\top}\dot{\mathbf{X}}$\upcite{RN212,RN209} rather than by explicitly computing the matrix inverse, where $\dagger$ denotes the pseudo-inverse. However, as the driving components substantially increase the scale of the feature matrix, the numerical errors arising from the pseudo-inverse approximation of the inverse matrix may become significant\upcite{RN259}. The large numerical errors can lead to substantial deviations between the learned dynamics and the true dynamics, since it causes the coefficient matrix $\mathbf{A}$ to deviate significantly from the true solution. This fact has been consistently verified through our experiments on various systems. In particular, this inaccuracy of inferring dynamics becomes pronounced when using a large basis function library which is helpful for capturing dynamics in networked systems\upcite{RN243,RN246}. Therefore, reducing the numerical errors introduced by the pseudo-inverse is fundamental for inferring reliable dynamical equations.

\subsection{Grid search for optimal driving variable}
Since the variable $\nu$ is represented as $\nu_{i+1}=\nu_i+s_i\Delta \nu$, it has two hyperparameters: the initial value $\nu_1$ and the step size $\Delta \nu$. To identify the optimal driving variable $\nu$ for constructing the basis functions, we perform a grid search over $11\times12$ candidate value combinations to find the optimal $\nu_1$ and $\Delta \nu$, which are $[0,\pm1,\pm5,\pm10,\pm15,\pm20]\times [1\times10^{-5}, 5\times10^{-5}, \dots, 1, 5]$ (see Fig. 2). The model performance metric for each pair $(\nu_1,\Delta \nu)$ is evaluated using a numerical error adjusted Akaike's information criterion (i.e., $\varepsilon$AIC), which is derived from the Akaike's information criterion\upcite{RN281} (AIC; $\text{AIC}=n\text{logMSE}+2p$). The $\varepsilon$AIC balances accuracy (including both numerical and fitting accuracy) and model complexity, which is defined as
\begin{equation*}
   \varepsilon\text{AIC} =
  \begin{cases}
    \varepsilon n\text{logMSE}+2p & \text{if } \text{logMSE} \geq 0, \\
    \frac{1}{\varepsilon}n\text{logMSE}+2p  & \text{if } \text{logMSE} < 0,
  \end{cases}
\end{equation*}
where $n$ is the number of observations, MSE is the mean squared error of the regression result of the model, $p$ is the number of variables, and $\varepsilon$ is the normalized numerical error introduced by pseudo-inverse approximation. Since $\varepsilon$ measures the numerical inaccuracy of the inference process, a model with a larger $\varepsilon$ is more likely to be unreliable. Thus, multiplying $\varepsilon$ or $1/\varepsilon$ by $\log\text{MSE}$ amplifies the impact of the fitting inaccuracy. The value of $\varepsilon$ is computed as
\begin{equation*}
    \varepsilon=\frac{\|(\Theta_{\nu}^{\top}\Theta_{\nu}+\alpha\mathbf{I})(\Theta_{\nu}^{\top}\Theta_{\nu}+\alpha\mathbf{I})^{\dagger}-\mathbf{I}\|_F^2}{N^2},
\end{equation*}
where $\|\cdot\|_F$ is the Frobenius norm, $\alpha$ is the regularization parameter ($\alpha=0$ for the ordinary least-squares case), $\dagger$ refers to the pseudo-inverse, $N^2$ is the number of elements in matrix $\Theta_{\nu}^{\top}\Theta_{\nu}$, and $\mathbf{I}$ is the identity matrix. The pair $(\nu_1,\Delta \nu)$ with the least $\varepsilon$AIC value is used to construct the optimal driving variable.

\subsection{Performance measures}
The performance measures used to evaluate the methods in this work are as follows:

The inaccuracy of the inferred equations for synthetic systems is quantified by symmetric mean absolute percentage error (sMAPE)\upcite{RN282}, which is
\begin{equation*}
    \text{sMAPE}=\frac{1}{m}\sum\limits_{i=1}^{m}\frac{\left| I_i(t)-R_i(t) \right|}{\left| I_i(t) \right|+\left| R_i(t) \right|},
\end{equation*}
where $m$ is the cardinal number of the set which contains inferred and true function terms, while $I_i(t)$ and $R_i(t)$ are the inferred and true coefficients for each term at time point $t$, respectively. The more accurate the inferred equations, the lower the value of sMAPE, which ranges from 0 to 1. This indicator captures not only the errors in inferred coefficients but also the incorrectness of the inferred equations form\upcite{RN243,RN246}.

The inaccuracy of the predicted dynamics is quantified using normalized Euclidean distance (NED), which represents the distance between the two trajectories generated by the inferred and true dynamical equations. That is,
\begin{equation*}
\text{NED}=\frac{\sqrt{\sum\limits_{i=1}^{d}\sum\limits_{t=t_0}^{T}(x_i(t)-\hat{x}_i(t))^2}}{\sqrt{\sum\limits_{i=1}^{d}\sum\limits_{t=t_0}^{T}(x_i^2(t)+\hat{x}_i^2(t))}},
\end{equation*}
where $d$ is the dimensionality of the study system, $x_i$ is the true trajectory of the $i$-th dimension, $\hat{x}_i$ is the trajectory of the $i$-th dimension generated by the inferred equations, and $t_0$ and $T$ are the beginning and ending times of the trajectory.

\section{Data availability}
All the synthetic and empirical data used in this paper are publicly available at the Github (\url{https://github.com/zhugchzo/infer_nonstation_dynamics}).

\section{Code availability}
All the source codes are publicly available at the Github (\url{https://github.com/zhugchzo/infer_nonstation_dynamics}).

\section{Acknowledgments}
This work was funded by the National Natural Science Foundation of China under grant 62388101 and the National Key Research and Development Program of China under grant 2022YFF0902800. The authors are also grateful for the helpful discussion with Weifeng Shang.

\section{Author contributions}
W.C. conceived the research. C.Z., Z.J., and W.C. designed the research. C.Z. analyzed the empirical data and did the analytical and numerical calculations. C.Z., Z.J., and W.C. analyzed the results. All authors wrote the manuscript.

\section{Competing interests}
The authors declare no competing interests.

\bibliographystyle{naturemag}
\bibliography{reference.bib}

\end{document}